\documentclass[fleqn,twoside]{article}
\usepackage{espcrc2}

\title{Exponential suppression with four legs
and an infinity of loops}

\author{David J. Broadhurst\address{Physics and Astronomy Department,
Open University, Milton Keynes MK7 6AA, UK}\thanks{Talk presented 
at {\em ``Loops and Legs in Quantum Field Theory"}, W{\"o}rlitz, Germany, 
April 2010}
and 
Andrei I. Davydychev\address{Institute for Nuclear Physics,
Moscow State University, 
119992 Moscow, Russia}%
\address{Schlumberger, HFE,
110 Schlumberger Dr., Sugar Land, TX 77478, USA}}

\begin{document}

\begin{abstract}
The $L$-loop 4-point ladder diagram of massless $\phi^3$ theory is finite
when all 4 legs are off-shell and is given in terms of polylogarithms
with orders ranging from $L$ to $2L$. We obtain the exact solution
of the linear Dyson--Schwinger equation that sums these ladder diagrams
and show that this sum vanishes exponentially fast at strong coupling.
\vspace{1pc}
\end{abstract}

\maketitle

\section{INTRODUCTION}

Results for two-loop 3-point and 4-point ladder diagrams 
shown in Figure~1 were
obtained in~\cite{UD1}. In~\cite{UD2},
results were found for an arbitrary number of loops, $L$.
These were confirmed in~\cite{DJB}, using
Gegenbauer-polynomial methods.
Here we shall sum the 4-point ladders of Figure~1b.


\newcommand{\threepoint}{
\setlength {\unitlength}{0.43mm}
\begin{picture}(152,75)(0,0)
\put (0,35) {\line(1,0){16}}
\put (16,35) {\line(4,1){131}}
\put (16,35) {\line(4,-1){131}}
\put (40,29) {\line(0,1){12}}
\put (64,23) {\line(0,1){24}}
\put (112,11) {\line(0,1){48}}
\put (136,5) {\line(0,1){60}}
\put (6,38) {\vector(1,0){9}}
\put (0,43) {\makebox(0,0)[bl]{\large $p_3$}}
\put (146,70) {\vector(-4,-1){9}}
\put (149,69) {\makebox(0,0)[bl]{\large $p_1$}}
\put (146,0) {\vector(-4,1){9}}
\put (149,0) {\makebox(0,0)[bl]{\large $p_2$}}
\put (77,35) {\makebox(0,0)[bl]{\huge $. \; . \; .$}}
\put (75,0){\makebox(0,0)[c]{\large ${\mbox{(a)}}$}}
\end{picture}
}
\newcommand{\fourpoint}{
\setlength {\unitlength}{0.43mm}
\begin{picture}(152,50)(0,0)
\put (5,10) {\line(1,0){143}}
\put (5,40) {\line(1,0){143}}
\put (18,10) {\line(0,1){30}}
\put (44,10) {\line(0,1){30}}
\put (70,10) {\line(0,1){30}}
\put (109,10) {\line(0,1){30}}
\put (135,10) {\line(0,1){30}}
\put (7,8) {\vector(1,0){8}}
\put (0,0) {\makebox(0,0)[bl]{\large $k_1$}}
\put (7,42) {\vector(1,0){8}}
\put (0,47) {\makebox(0,0)[bl]{\large $k_2$}}
\put (145,42) {\vector(-1,0){8}}
\put (149,47) {\makebox(0,0)[bl]{\large $k_3$}}
\put (145,8) {\vector(-1,0){8}}
\put (149,0) {\makebox(0,0)[bl]{\large $k_4$}}
\put (78,25) {\makebox(0,0)[bl]{\huge $.\; .\; .$}}
\put (75,0){\makebox(0,0)[c]{\large ${\mbox{(b)}}$}}
\end{picture}
}
\begin{figure}[bth]
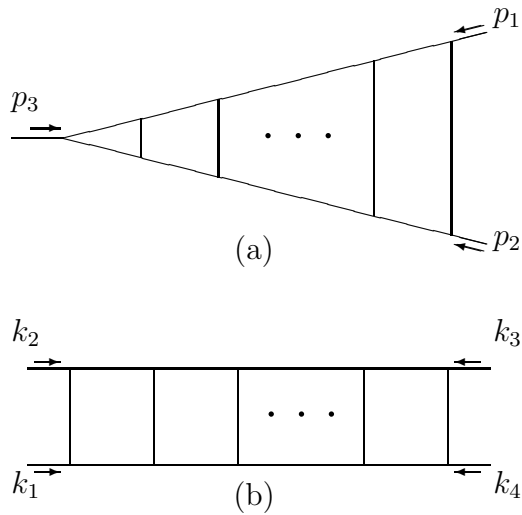

\[
\begin{array}{c}
\threepoint \\[10mm] 
\fourpoint
\end{array}
\]
\caption{(a) 3-point, and (b) 4-point, $L$-loop diagrams in $\phi^3$ theory}
\end{figure}


With massless internal propagators,
each diagram gives a finite real contribution for positive values
of the 6 kinematic invariants 
\[
k_1^2,~k_2^2,~k_3^2,~k_4^2,~s=(k_1+k_2)^2,~t=(k_2+k_3)^2.
\]
We shall investigate
the strong-coupling limit of the exact solution
to a linear Dyson--Schwinger equation, of the schematic form 
\begin{equation}
\label{SD_BS}
{\cal D}={\cal T}+g^2\int{\rm d}^4k\;{\cal T}\cdot{\cal D},
\end{equation} 
which sums these ladder diagrams of Figure~1b. 
Here, ${\cal T}$ is the $t$-channel tree-diagram, which we normalize to $1/t$,
and the dot indicates convolution under the 4-dimensional integration
that adds another loop (including the factor ${\rm i}(2\pi)^{-4}$). 
Note that the function ${\cal D}$ 
in Equation~(\ref{SD_BS}) 
can be also understood as the kernel of Bethe--Salpeter equation
in the ladder approximation (see also in Ref.~\cite{Arbuzov}).

We offer two motivations for this investigation.
First, it was shown in~\cite{DJB} that
the ladder diagrams contributing to the derivative of the self-energy
energy of massless $\phi^3$ theory sum to give the constant
$-\frac12\zeta(-1)=\frac{1}{24}$ at infinite coupling.
As far back as 1993, we suspected that by including
the tree-diagram ${\cal T}$ in ${\cal D}$ we would obtain zero for the
sum of 4-point ladder diagrams at infinite coupling.
It has taken us 17 years to prove that this indeed is the case.
Secondly, and more recently,
we have noted that ladder approximations are of interest
to workers in ${\cal N}=4$ super Yang--Mills 
theory \cite{N=4sym,Drummond}, 
whose strong coupling
limit may be governed by an AdS/CFT correspondence.
Another interesting application is the conformal quantum
mechanics~\cite{Isaev}. We also note that some properties
of the functions occurring in~\cite{UD1,UD2} were studied 
in~\cite{Kondrashuk}. Certain asymptotic limits of ladder diagrams
in $\phi^3$ theory were studied in~\cite{Osland_Wu}.

In any case, we hope
that it may be of interest to colleagues to see
the explicit form of a 4-point ladder sum, as a function of the
6 kinematic invariants and the coupling $g^2$, which also
has the dimensions of (mass)$^2$ in $\phi^3$ theory.

It was a pleasant surprise to us to obtain,
eventually, the solution to this toy problem
as a single integral of elementary
functions that manifestly vanishes exponentially fast
as the dimensionless coupling $g^2/(4\pi^2s)$
tends to infinity. The polylogarithmic complexity of perturbation
theory is in marked contrast to the simplicity of the all-orders
result that we shall now derive.

\section{THE $L$-LOOP TERM}

We write the perturbation series of ladder diagrams as
\begin{eqnarray}
\label{calD}
&&\hspace*{-7mm}
{\cal{D}}(k_1^2,k_2^2,k_3^2,k_4^2,s,t)
\nonumber \\
&& = \frac{1}{t}
\left\{ 1+ \sum\limits_{L=1}^{\infty}\;
\left( - \frac{\kappa^2}{4} \right)^L
\Phi^{(L)}(X,Y) \right\}
\end{eqnarray}
with dimensionless ratios
\begin{equation}
\label{XY}
X \equiv \frac{k_1^2 k_3^2}{s t} ,\quad
Y \equiv \frac{k_2^2 k_4^2}{s t} ,\quad
\kappa^2\equiv\frac{g^2}{4\pi^2s}
\end{equation}
that we assume to be positive. As shown in~\cite{UD2}, the $L$-loop term
\begin{eqnarray}
\label{PhiLint}
&&\hspace*{-7mm}
\Phi^{(L)} (X,Y) = - \frac{1}{L! \; (L-1)!}
\nonumber \\
&& \times\int_0^1
\frac{{\rm d} \xi}{Y \xi^2 + (1-X-Y) \xi +X} \; \hspace{2cm}
\nonumber \\
&& \times\left[\ln \xi \left(\ln \frac{Y}{X} \!+\! \ln{\xi} 
\right)\right]^{L-1}
\left( \ln \frac{Y}{X}\! + \!2\ln{\xi} \right)
\end{eqnarray}
depends only on the cross ratios $X$ and $Y$ and is described 
by the same function as the ladder 3-point function. 
The origin of this
simplification was elucidated in~\cite{DJB}, which gave the conformal
transformation that relates Figure~1b to Figure~1a. When scaled by
an appropriate power of $p_3^2$, the latter depends only
on the ratios $x=p_1^2/p_3^2$ and $y=p_2^2/p_3^2$ and is
given by $\Phi^{(L)} (x,y)$.

The integral (\ref{PhiLint}) may be evaluated in terms of
polylogarithms~\cite{UD2}. Here, we shall consider the case where the K\"{a}llen function
\begin{equation}
\label{def_mu}
\mu=\sqrt{4XY-(X+Y-1)^2}
\end{equation}
is real and positive. Then we are comfortably outside the region that contains
Landau singularities and hence may define the geometrical angle~\cite{phi3}
(see also in~\cite{geom})
\begin{equation}
\label{phi}
\phi = \arccos\left(\frac{X+Y-1}{\sqrt{4XY}}\right)
\end{equation}
with $\pi>\phi>0$. In this region, the $L$-loop term~\cite{UD2}
\begin{eqnarray}
\label{Phi_L_3b}
\Phi^{(L)}(X,Y) &\!\!=\!\!& \frac{2}{\mu L!}
\sum\limits_{j=L}^{2L}
\frac{j!}{(j\!-\!L)!\; (2L\!-\!j)!}
\nonumber \\
&& \hspace*{-5mm} \times
\left(\ln\frac{X}{Y}\right)^{2L-j}
\rm{Im}\;{\rm Li}_j\left(\sqrt{\frac{Y}{X}}\;e^{{\rm i}\phi}\right)
\end{eqnarray}
is given in terms of products of powers $\ell\equiv\ln(X/Y)$
and the imaginary parts of polylogarithms ${\rm Li}_j$
with orders running from $j=L$ to $j=2L$.
The symmetry $\Phi^{(L)}(X,Y)=\Phi^{(L)}(Y,X)$ is ensured
by the inversion formula for polylogarithms, given in~\cite{Lewin}.

\section{INFINITE SUM OF LADDER DIAGRAMS}

\subsection{An integral with a Bessel function}

\vspace*{3mm}

In the first instance we omit the tree term and use
the integral representation~(\ref{PhiLint}) to sum the series
\begin{eqnarray}
\label{J_1-XY}
&&\hspace*{-7mm}
\sum\limits_{L=1}^{\infty}\;
\left( - \frac{\kappa^2}{4} \right)^L
\Phi^{(L)}(X,Y)
\nonumber \\
&& = \frac{\kappa}{2} \int\limits_0^1
\frac{\mbox{d}\xi}{X+(1-X-Y)\xi+Y\xi^2}\;
\nonumber\\
&& \quad \times
\left( \ln\frac{Y}{X} + 2\ln\xi \right)
\frac{1}{\sqrt{\ln\xi \left(\ln\frac{Y}{X}+\ln\xi\right) }} \;
\nonumber\\
&& \quad \times
J_1\left( \kappa \sqrt{\ln\xi \left(\ln\frac{Y}{X}+\ln\xi\right) } \right)
\end{eqnarray}
where $J_1$ is a Bessel function.
Substituting $\xi=e^{-\eta}$ and denoting
$\ell\equiv \ln\frac{X}{Y}$,
we obtain
\begin{eqnarray}
\label{PhiL_2}
&&\hspace*{-7mm}
\sum\limits_{L=1}^{\infty}\;
\left( - \frac{\kappa^2}{4} \right)^L
\Phi^{(L)}(X,Y)
\nonumber \\
&& = -\frac{\kappa}{2}
\int\limits_0^{\infty}
\frac{e^{-\eta}\;{\rm d}\eta}
{X + (1-X-Y) e^{-\eta}+Y e^{-2\eta}}
\nonumber \\ 
&& \quad \times
\frac{2\eta+\ell}{\sqrt{\eta(\ell+\eta)}}
J_1\left(\kappa\sqrt{\eta(\ell+\eta)}\right).
\end{eqnarray}

The denominator in (\ref{PhiL_2}) may be re-written as
\begin{eqnarray}
&&\hspace*{-7mm}
X + (1-X-Y) e^{-\eta}+Y e^{-2\eta}
\nonumber \\
&& = e^{-\eta} \left[ 1-X-Y
+ 2\sqrt{XY} \cosh\!\left(\eta+\frac{\ell}{2}\right)\right]
\nonumber \\
&& = -2\sqrt{XY} e^{-\eta}
\left[ \cos\phi - \cosh\!\left(\eta+\frac{\ell}{2}\right)\right].
\end{eqnarray}
In this way, we arrived at
\begin{eqnarray}
\label{PhiL_3}
&&\hspace*{-7mm}
\sum\limits_{L=1}^{\infty}\;
\left( - \frac{\kappa^2}{4} \right)^L
\Phi^{(L)}(X,Y)
\nonumber \\
&&= \frac{\kappa}{4\sqrt{XY}}
\int\limits_0^{\infty}
\frac{{\rm d}\eta}
{\cos\phi - \cosh\!\left(\eta+\frac{\ell}{2}\right)}
\nonumber \\
&& \quad \times \frac{2\eta+\ell}{\sqrt{\eta(\ell+\eta)}}
J_1\left(\kappa\sqrt{\eta(\ell+\eta)}\right)
\end{eqnarray}
and obtained, in 1993, an explicit summation
of all 4-point ladder diagrams with loop numbers $L>0$.
Yet we could find no way of investigating our hunch that
inclusion of the tree diagram, with $L=0$, might give
an exponentially vanishing result at infinitely strong coupling.

The first break-through came from noticing that
\begin{eqnarray}
\label{J1_J0_diff}
&&\hspace*{-7mm}
\frac{2\eta+\ell}{\sqrt{\eta(\ell+\eta)}}\;
J_1\left(\kappa\sqrt{\eta(\ell+\eta)}\right)
\nonumber \\
&& = -\frac{2}{\kappa}\;
\frac{{\rm d}}{{\rm d}\eta}
J_0\left(\kappa\sqrt{\eta(\ell+\eta)}\right).
\end{eqnarray}
Then, integrating by parts, we found that the
Dyson--Schwinger solution is
\begin{eqnarray}
\label{calD_2}
&&\hspace*{-7mm}
{\cal{D}}(k_1^2,k_2^2,k_3^2,k_4^2,s,t)
\nonumber \\
&& \hspace*{-7mm}
= \frac{1}{2t\sqrt{XY}}
\int\limits_0^{\infty} {\rm d}\eta\;
\frac{\sinh\left(\eta\!+\!\frac{\ell}{2}\right)
 J_0\!\left(\kappa\sqrt{\eta(\ell\!+\!\eta)}\right)}
 {\left[\cosh\!\left(\eta+\frac{\ell}{2}\right)-\cos\phi\right]^2}
\end{eqnarray}
where the tree-term $1/t$ is precisely included by the surface term
of the partial integration enabled by~(\ref{J1_J0_diff}).
Our hopes had increased: the full Dyson--Schwinger 
solution~(\ref{calD_2}) is now presented in a form that looks more
promising for confirmation of our guess of exponential suppression at strong coupling.

Next, we shift the integration variable $\eta$ and obtain
\begin{eqnarray}
\label{calD_3}
&&\hspace*{-7mm}
{\cal{D}}(k_1^2,k_2^2,k_3^2,k_4^2,s,t)
\nonumber \\
&&\hspace*{-3mm} 
= \frac{1}{2t\sqrt{XY}}
\int\limits_{\ell/2}^{\infty} {\rm d}\eta\;
\frac{\sinh\eta\;
 J_0\!\left(\kappa\sqrt{\eta^2-\textstyle{1\over4}\ell^2}\right)}
 {\left(\cosh\eta-\cos\phi\right)^2}.
\end{eqnarray}
The $X\longleftrightarrow Y$ symmetry of
expression~(\ref{calD_3}) is now quite easy to understand.
If we were to interchange $X$ and $Y$, then
the only thing that would change is the lower limit of integration:
$\ell/2\rightarrow -\ell/2$,
since $\phi\equiv\arccos((X+Y-1)/\sqrt{4XY})$ is symmetric in $(X,Y)$.
The integral between
$-\ell/2$ and $\ell/2$ is zero, since the integrand is an odd function
of $\eta$ and an even function of $\ell\equiv\ln(X/Y)$.
Hence we may take $\frac12|\ell|=\frac12|\ln X-\ln Y|$ as the lower limit
of integration in~(\ref{calD_3}).

\subsection{Exponential suppression at strong coupling}

\vspace*{3mm}

We re-write~(\ref{calD_3}) as
\begin{eqnarray}
\label{calD_4}
&&\hspace*{-7mm}
{\cal{D}}(k_1^2,k_2^2,k_3^2,k_4^2,s,t)
\nonumber \\
&& = \frac{1}{2t\sqrt{XY}}
\int\limits_{0}^{\infty}
\frac{{\rm d}\eta\; \sinh\eta}
 {\left(\cosh\eta-\cos\phi\right)^2}\;
\nonumber \\
&& \quad \times
J_0\!\left(\kappa\sqrt{\eta^2-\textstyle{1\over4}\ell^2}\right)
\vartheta\left( \eta^2-\textstyle{1\over4}\ell^2 \right) ,
\end{eqnarray}
where $\vartheta(x)$ is the Heaviside function, with $\vartheta(x)=1$, for
$x>0$, and $\vartheta(x)=0$, otherwise.
Now, let us use the integral representation
\begin{eqnarray}
\label{integral_J0_theta}
&&\hspace*{-7mm}
\int\limits_0^{\infty} {\rm d}\tau\; \sin\left(\kappa\eta\cosh\tau\right)\;
\cos\left({\textstyle{\frac{1}{2}\ell\kappa\sinh\tau}}\right)
\nonumber \\
&&= \frac{\pi}{2}\; J_0\!\left(\kappa\sqrt{\eta^2-\textstyle{1\over4}\ell^2}\right)\;
\vartheta\left( \eta^2-\textstyle{1\over4}\ell^2 \right)
\end{eqnarray}
which may be obtained from Equation~(2.5.25.9) of~\cite{PBM1}
(with the substitutions $x=\kappa\sinh\tau$, $y=\kappa$,
$c=\eta$, and $b=\frac{1}{2}\ell$).
The key point is that
we are rid of the integration limit $\ell/2$.

By this device, we obtain
\begin{eqnarray}
&&\hspace*{-7mm}
{\cal{D}}(k_1^2,k_2^2,k_3^2,k_4^2,s,t)
\nonumber \\
&&= \frac{1}{\pi t\sqrt{XY}}
\int\limits_{0}^{\infty}
\frac{{\rm d}\eta\; \sinh\eta}
 {\left(\cosh\eta-\cos\phi\right)^2}\;
\nonumber \\
&&\quad \times 
\int\limits_0^{\infty} {\rm d}\tau \sin\left(\kappa\eta\cosh\tau\right)
\cos\left({\textstyle{\frac{1}{2}}}\ell\kappa\sinh\tau\right)
\end{eqnarray}
as a double integral.
Next, the substitution $z=\kappa\cosh\tau$
gives $\kappa\sinh\tau=\sqrt{z^2-\kappa^2}$ and
${\rm d}\tau={\rm d}z/\sqrt{z^2-\kappa^2}$. Hence we obtain
\begin{eqnarray}
&&\hspace*{-7mm}
{\cal{D}}(k_1^2,k_2^2,k_3^2,k_4^2,s,t)
\nonumber \\
&=& \frac{1}{\pi t\sqrt{XY}}
\int\limits_{0}^{\infty}
\frac{{\rm d}\eta\; \sinh\eta}
 {\left(\cosh\eta-\cos\phi\right)^2}\;
\nonumber \\
&& \times \int\limits_{\kappa}^{\infty} \frac{{\rm d}z\; 
\sin(\eta z)}{\sqrt{z^2-\kappa^2}}\;
\cos\left( {\textstyle{\frac{1}{2}}} \ell\sqrt{z^2-\kappa^2}\right).
\end{eqnarray}

Now we reverse the order of the integrations, obtaining
\begin{eqnarray}
&&\hspace*{-7mm}
{\cal{D}}(k_1^2,k_2^2,k_3^2,k_4^2,s,t)
\nonumber \\
&=& \frac{1}{\pi t\sqrt{XY}}
\int\limits_{\kappa}^{\infty} \frac{{\rm d}z}{\sqrt{z^2-\kappa^2}}\;
\cos\left( {\textstyle{\frac{1}{2}}} \ell\sqrt{z^2-\kappa^2}\right)
\nonumber \\
&& \times \int\limits_{0}^{\infty}
\frac{{\rm d}\eta\; \sinh\eta\; \sin(\eta z)}
 {\left(\cosh\eta-\cos\phi\right)^2}\; .
\end{eqnarray}
From Equation~(2.5.48.18) of~\cite{PBM1} (with $t=\pi-\phi$, $c=1$, $b=z$),
we obtain
\begin{equation}
\label{PBM_1}
\int\limits_{0}^{\infty}
\frac{{\rm d}\eta\; \sinh\eta\; \sin(\eta z)}
 {\left(\cosh\eta-\cos\phi\right)^2}
=\frac{\pi z}{\sin\phi}
\;\frac{\sinh\left[(\pi-\phi)z\right]}{\sinh(\pi z)} .
\end{equation}

Recalling that $\mu=2\sqrt{XY} \sin\phi$, we obtain
\begin{eqnarray}
\label{AP}
&&\hspace*{-7mm}
{\cal{D}}(k_1^2,k_2^2,k_3^2,k_4^2,s,t)
\nonumber \\
&=& \frac{2}{t\mu}
\int\limits_{\kappa}^{\infty} \frac{z\; {\rm d}z}{\sqrt{z^2-\kappa^2}}\;
\frac{\sinh\left[(\pi-\phi)z\right]}{\sinh(\pi z)}\;
\nonumber \\
&& \times
\cos\left( {\textstyle{\frac{1}{2}}} \ell\sqrt{z^2-\kappa^2}\right) \; .
\end{eqnarray}
This is our final solution to the Dyson--Schwinger equation~(\ref{SD_BS}) that
sums all $L$-loop 4-point ladder diagrams, including (most crucially)
the tree-diagram, with $L=0$ loops.
The sum manifestly vanishes, exponentially fast, as the dimensionless
coupling $\kappa=g/(2\pi\sqrt{s})$ tends to infinity, since
the ratio of sinh functions in the integrand of~(\ref{AP}) satisfies
\begin{equation}
\frac{\sinh\left[(\pi-\phi)z\right]}{\sinh(\pi z)}\le
\frac{\sinh\left[(\pi-\phi)\kappa\right]}{\sinh(\pi\kappa)}=
{\cal O}(e^{-\kappa\phi})
\end{equation}
with $\pi>\phi>0$. 

So we are done, 17 years after conjecturing such an exponential suppression.

\section{COMMENTS}

Our actual route to this final answer bears scant relation 
to the more coherent explanation offered here. After many 
fruitful exchanges of ideas, the first author guessed the 
final result, by means far too involved to be recounted 
here, and then the second author neatly devised a process of 
reverse-engineering that resulted in the proof presented 
here, via formulae presented in~\cite{PBM1,PBM2}.

It is not clear to either of us whether our explicit all-orders summation
of 4-point ladder diagrams may still hold some interest for the 
loops-and-legs community that has nurtured our efforts. Yet we hope that it might.
In any case, it was fun to achieve. 

\vspace*{8mm}

We gratefully acknowledge the crucial role of Natalia Ussyukina and the
moral support of Bas Tausk and Dirk Kreimer, which sustained our resolve.


\begin{thebibliography}{99}

\bibitem{UD1} 
N.I.~Ussyukina and A.I.~Davydychev,
  Phys.\ Lett.\  B298 (1993) 363.

\bibitem{UD2} 
N.I.~Ussyukina and A.I.~Davydychev,
  Phys.\ Lett.\  B305 (1993) 136.

\bibitem{DJB} 
D.J.~Broadhurst,
  Phys.\ Lett.\  B307 (1993) 132.

\bibitem{Arbuzov}
B.A.~Arbuzov and V.E.~Rochev,
  Sov.\ J.\ Nucl.\ Phys.\  21 (1975) 455;
K.G.~Klimenko and V.E.~Rochev,
  Theor.\ Math.\ Phys.\  32 (1978) 787.

\bibitem{N=4sym}
B.~Eden, P.S.~Howe, C.~Schubert, E.~Sokatchev and P.C.~West,
  Nucl.\ Phys.\ B557 (1999) 355;
B.~Eden, C.~Schubert and E.~Sokatchev,
  Phys.\ Lett.\  B482 (2000) 309;
M.~Bianchi, S.~Kovacs, G.~Rossi and Y.S.~Stanev,
  Nucl.\ Phys.\  B584 (2000) 216;
F.A.~Dolan and H.~Osborn,
  Nucl.\ Phys.\  B599 (2001) 459;
N.~Beisert, C.~Kristjansen, J.~Plefka, G.W.~Semenoff and M.~Staudacher,
  Nucl.\ Phys.\  B650 (2003) 125;
J.M.~Drummond, G.P.~Korchemsky and E.~Sokatchev,
  Nucl.\ Phys.\  B795 (2008) 385;
D.~Nguyen, M.~Spradlin and A.~Volovich,
  Phys.\ Rev.\  D77 (2008) 025018;
L.F.~Alday and R.~Roiban,
  Phys.\ Rept.\  468 (2008) 153;
G.C.~Rossi and Y.S.~Stanev,
  Nucl.\ Phys.\  B807 (2009) 534;
B.~Basso and G.P.~Korchemsky,
  J.\ Phys.\ A42 (2009) 254005.

\bibitem{Drummond}
J.M.~Drummond, J.~Henn, V.A.~Smirnov and E.~Sokatchev,
  JHEP 0701 (2007) 064.

\bibitem{Isaev}
A.P.~Isaev,
  Nucl.\ Phys.\  B662 (2003) 461;
A.P.~Isaev,
  Phys.\ Atom.\ Nucl.\  71 (2008) 914.
  
\bibitem{Kondrashuk}
I.~Kondrashuk and A.~Kotikov,
  JHEP 0808 (2008) 106;
I.~Kondrashuk and A.~Vergara,
  JHEP 1003 (2010) 051.
  
\bibitem{Osland_Wu}
P.~Osland and T.T.~Wu,
  Nucl.\ Phys.\  B288 (1987) 77, 95.

\bibitem{phi3}
A.I.~Davydychev and J.B.~Tausk,
  Nucl.\ Phys.\  B397 (1993) 123;
  Phys.\ Rev.\  D53 (1996) 7381;
A.I.~Davydychev,
  Phys.\ Rev.\  D61 (2000) 087701.
  
\bibitem{geom}
A.I.~Davydychev and R.~Delbourgo,
  J.\ Math.\ Phys.\  39 (1998) 4299;
A.I.~Davydychev and M.Yu.~Kalmykov,
  Nucl.\ Phys.\  B605 (2001) 266.

\bibitem{PBM1} 
A.P.~Prudnikov, Yu.A.~Brychkov and O.I.~Marichev,
  Integrals and Series, Vol.~1, Nauka, Moscow, 1981.

\bibitem{PBM2} 
A.P.~Prudnikov, Yu.A.~Brychkov and O.I.~Marichev,
  Integrals and Series, Vol.~2, Nauka, Moscow, 1983.

\bibitem{Lewin} 
L.~Lewin, Polylogarithms and Associated Functions,
  North-Holland, Amsterdam, 1981.

\end{thebibliography}
\end{document}